# From Likelihood to Limit State: A Reliability-Inspired Framework for Bayesian Evidence Estimation and High-dimensional Sampling

Zihan Liao[a,b], Binbin Li[a,b], Hua-Ping Wan[b*]

[a] State Key Laboratory of Biobased Transportation Fuel Technology, ZJU-UIUC Institute, Zhejiang University, Haining, Zhejiang, China

[b] College of Civil Engineering and Architecture, Zhejiang University, Hangzhou, Zhejiang China

[*] Corresponding author; E-mail address: hpwan@zju.edu.cn

# Abstract

Bayesian analysis plays a crucial role in estimating distribution of unknown parameters for given data and model. Due to the curse of dimensionality, it becomes difficult for high-dimensional problems, especially when multiple modes exist. This paper introduces an efficient Bayesian posterior sampling algorithm, based on a new interpretation of evidence from the perspective of structural reliability estimation. That is, the evidence can be equivalently formulated as an integration of failure probabilities, by regarding the likelihood function as a limit state function. The evidence is then evaluated with subset simulation (SuS) algorithm. The posterior samples can be obtained following the principle of importance resampling as a postprocessing procedure. The estimation variance is derived to quantify the inherent uncertainty associated with the SuS estimator of evidence. The effective sample size is introduced to measure the quality of posterior sampling. Three benchmark examples are first considered to illustrate the performance of the proposed algorithm by comparing it with two state-of-art algorithms. It is then used for the finite element model updating, showing its applicability in practical engineering problems. The proposed SuS algorithm exhibits comparable or even better performance in evidence estimation and posterior sampling, compared to the aBUS and MULTINEST algorithms, especially when the dimension of unknown parameters is high.

## 1. Introduction

Bayesian analysis is the modern engine of data science [1] and has a wide application in civil engineering [2,3]. The main objective of Bayesian analysis is to obtain the posterior distribution of unknown parameters, which incorporates the prior knowledge and all the information from the data. Consider a model $\mathcal{M}(\boldsymbol{\theta})$, e.g., a finite element (FE) model of a structure, consisting of unknown parameters $\boldsymbol{\theta} \in \mathbb{R}^d$. The Bayesian approach regards $\boldsymbol{\theta}$ as a random variable (RV), with prior probability density function (PDF) $\pi(\boldsymbol{\theta})$, encoding the information from prior knowledge, e.g., engineering experience. We consider $\boldsymbol{\theta}$ as a continuous RV in this paper without loss of generality. Given measured data $\boldsymbol{D}$, the Bayes' theorem states that the posterior PDF $p(\boldsymbol{\theta}|\boldsymbol{D})$ can be obtained as

$$p(\boldsymbol{\theta}|\boldsymbol{D}) = \frac{1}{z}L(\boldsymbol{\theta};\boldsymbol{D})\pi(\boldsymbol{\theta}) \tag{1}$$

where the term $L(\boldsymbol{\theta};\boldsymbol{D})$ is the likelihood function, representing the information contained in data $\boldsymbol{D}$; and the normalizing constant

$$z = \int_\Omega L(\boldsymbol{\theta};\boldsymbol{D})\pi(\boldsymbol{\theta})d\boldsymbol{\theta} \tag{2}$$

is also called the 'marginal likelihood' or the 'Type-II likelihood' or the 'evidence'. The evidence plays a negligible role in the identification, but it is of critical importance in the model selection and averaging when there is a collection of competing models.

It is never an easy task to compute the posterior distribution and the evidence in the Bayesian analysis. One such complexity originates from the nonlinear, implicit nature of the likelihood function $L(\boldsymbol{\theta};\boldsymbol{D})$, combined with its high computational cost. For example, when applied in the engineering domain, this function often requires a time-consuming FE analysis. Note that the dependence of likelihood on data $\boldsymbol{D}$ will be neglected hereafter if no ambiguity arises. Furthermore, as the number of unknown parameters in a FE model becomes large, the well-known "curse of dimensionality" arises, compounding the difficulty in approximating the posterior PDF. With the increased dimensions, regions with high likelihood values become relatively small and isolated within the huge parameter space, and thus it becomes challenging to locate the posterior modes. These challenges have attracted significant scholarly interest, leading to the development of various methods, such as the Laplace approximation [4–6], the Chib's method [7,8], and the Monte Carlo (MC) sampling. Among these methods, the MC sampling has gained greater attention because they are guaranteed to converge to the correct posterior given enough samples.

Among MC sampling algorithms, annealing [9–11] and vertical likelihood representation [12–16] are two typical schemes proposed to transit random samples from prior PDF to posterior PDF through specially crafted proposal distributions [17]. Annealing methods introduce an "inverse temperature" parameter $\beta$ as the power of the likelihood to formulate a transitional path from the prior ($\beta = 0$) to the posterior ($\beta = 1$). According to detailed implementations, various methods have been developed, such as stepping-stone sampling [11], annealed importance sampling [9], power posteriors [10], and transitional Markov chain Monte Carlo (MCMC) [18]. The effectiveness of the annealing method highly depends on the selection of temperature settings. It often requires manual tuning, and the optimal strategy may be problem dependent [19].

Vertical likelihood MC regards the likelihood function $L(\boldsymbol{\theta})$ as an augmented RV and explores the likelihood space to aid the transition of samples. One typical method is the nested sampling (NS) [12], which iteratively contracts the prior volume and increases the likelihood threshold of samples until the desired precision is achieved. While notable advancements have been made in its various implementations, e.g., MULTINEST [13] and POLYCHORD [15], many challenges still exist. The core strategy of MULTINEST involves segmenting the sampling region through successive ellipses. However, potentially biased estimations can happen when the shape of the high likelihood area deviates from elliptical configuration. In POLYCHORD, the generation of samples is expected to be independent, but achieving this has been proven to be challenging [19]. A new development of vertical likelihood MC, known as Bayesian updating with structural reliability methods (BUS) [16], was proposed recently by converting the evidence evaluation into an equivalent reliability estimation problem. It is then solved by the subset simulation (SuS) algorithm [20]. An adaptive version, aBUS, was further proposed [21] to adaptively choose a key parameter $c$, which is defined to be the reciprocal of the upper bound of the likelihood function. Although its formulation is similar to the rejection sampling, there are apparent differences, e.g., the samples generated in BUS are correlated (even repeated). In addition, it conducts the evidence evaluation and posterior sampling with two different sets of samples, thus wasting the computation power.

Motivated by NS and BUS, an alternative method for Bayesian analysis based on SuS is proposed, incorporating a new interpretation of evidence as a failure probability. It takes advantage of SuS for high-dimensional and multi-modal sampling. Unlike BUS, it does not require tuning parameters, and all samples generated can directly participate in the Bayesian inference. It discards the independence assumption in NS, and the parallel MCMC setting allows a more efficient sampling from the posterior space. This paper is organized as follows. A new interpretation of evidence in Bayesian inference is provided from the perspective of reliability estimation in Section 2. The proposed algorithm is then detailed in Section 3. Its effectiveness is demonstrated via various examples and comparison with NS and aBUS in Section 4. Final conclusions are made in Section 5.

## 2. A New Interpretation of Evidence

The evidence in Bayesian inference is a multivariate integration, which is difficult to evaluate when the dimension $d$ is large. An analogous situation is encountered in the field of reliability estimation [22], which concerns the evaluation of the failure probability of a rare event. The connection between evidence and reliability estimation is explored in this section, which provides a new interpretation of evidence from the perspective of failure probability evaluation.

Given the joint PDF $\pi(\boldsymbol{\theta})$ of the model parameter $\boldsymbol{\theta} \in \mathbb{R}^d$, the failure event defined by $\Omega_f = \{\boldsymbol{\theta} \in \Omega: g(\boldsymbol{\theta}) > 0\}$ occurs with the probability of

$$p_f = \int_{\Omega_f} \pi(\boldsymbol{\theta}) d\boldsymbol{\theta} = \int_{\Omega} \mathbb{1}(g(\boldsymbol{\theta}) > 0)\pi(\boldsymbol{\theta}) d\boldsymbol{\theta} \qquad (3)$$

where $g(\boldsymbol{\theta})$ is called as the limit state function (LSF) in the field of reliability. It defines the boundary between the safe and failure domains. The indicator function "$\mathbb{1}(\cdot)$" equals one if $g(\boldsymbol{\theta}) > 0$ and zero otherwise. The failure probability in Eqn. (3) cannot be solved analytically in most cases because of the multivariate integration. Approximation methods have been proposed in the past decades, e.g., FORM/SORM by approximating the LSF and importance sampling/SuS using MC integration. A good introduction of these methods can be found in the textbook [22].

In Bayesian analysis, assuming that the likelihood $L(\boldsymbol{\theta})$ is upper bounded by $L_{\text{sup}} = \sup\{L(\boldsymbol{\theta}): \boldsymbol{\theta} \in \Omega\}$, one can equivalently write the evidence expression in Eqn. (2) as:

$$z = \int_\Omega \int_0^{L_{\sup}} \mathbb{1}(L(\boldsymbol{\theta}) > l) dl\, \pi(\boldsymbol{\theta}) d\boldsymbol{\theta} = \int_0^{L_{\sup}} \underbrace{\left[\int_\Omega \mathbb{1}(L(\boldsymbol{\theta}) > l) \pi(\boldsymbol{\theta}) d\boldsymbol{\theta}\right]}_{p_f(l)} dl \tag{4}$$

where we have used the trick $L(\boldsymbol{\theta}) = \int_0^{L(\boldsymbol{\theta})} dl = \int_0^{L_{\sup}} \mathbb{1}(L(\boldsymbol{\theta}) > l) dl$ for the first equation and assumed interchangeability of the integration with respect to (w.r.t.) $\boldsymbol{\theta}$ and $l$. The term in the bracket is defined as $p_f(l)$ and named as the failure probability function (FPF), because it represents the probability of the failure event with the LSF $g(\boldsymbol{\theta}) = L(\boldsymbol{\theta}) - l$ if $\boldsymbol{\theta} \sim \pi(\boldsymbol{\theta})$, when comparing it with Eqn. (3). From the definition, the FPF $p_f(l)$ is a nonincreasing function within the range of $[0,1]$ for $l \in [0, L_{\sup}]$. If one can compute $p_f(l)$ for every likelihood level $l$, the evidence evaluation can be reduced to a one-dimensional integration w.r.t. $l \in [0, L_{\sup}]$, as indicated by Eqn. (4).

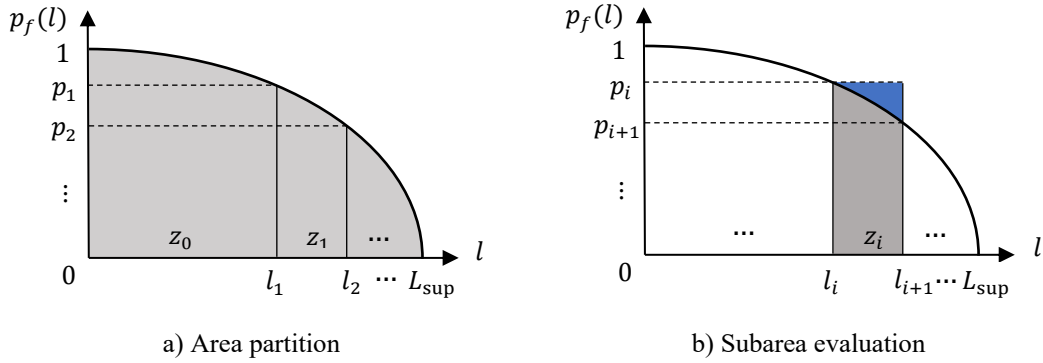

a) Area partition    b) Subarea evaluation

Figure 1. Evidence evaluation via FPF $p_f(l)$

One way to evaluate the evidence in Eqn. (4) is to compute the FPF $p_i = p_f(l_i)$, e.g., using the FORM or SORM method [22], and then obtain the result following the rule of numerical integration. However, this process is tedious and does not work well when the posterior distribution has multiple modes. In this paper, we propose a solution based on the principle of SuS. It partitions the range $[0, L_{\sup}]$ into a series of mutually exclusive and collectively exhaustive strata $[l_i, l_{i+1}]$ $(0 = l_0 < l_1 < \cdots < l_M = L_{\sup})$ and evaluates the partitioned subarea $z_i$ based on MC integration as illustrated in Figure 1. Suppose we have pairs $(l_i, p_i)$, one can compute $z_i$ as:

$$z_i = p_i(l_{i+1} - l_i) \int_\Omega \min\left\{\frac{L(\boldsymbol{\theta}) - l_i}{l_{i+1} - l_i}, 1\right\} \underbrace{\left[\frac{1}{p_i}\mathbb{1}(L(\boldsymbol{\theta}) > l_i)\pi(\boldsymbol{\theta})\right]}_{q(\boldsymbol{\theta}|l_i)} d\boldsymbol{\theta} \tag{5}$$

which can be understood as the rectangular area $p_i(l_{i+1} - l_i)$ multiplied by a fraction given by the integration term in Eqn. (5). The fraction can be evaluated by MC from the conditional PDF $q(\boldsymbol{\theta}|l_i)$.

For the illustration purpose, we choose to work with the coordinate of likelihood function $L(\boldsymbol{\theta})$. However, it is more beneficial to work with the log likelihood function $\mathcal{L}(\boldsymbol{\theta}) = \ln L(\boldsymbol{\theta})$ for its numerical stability. Defining $\ell_i = \ln l_i$ and $f_i(\boldsymbol{\theta}) = \exp(\ell_i) \min\{\exp(\mathcal{L}(\boldsymbol{\theta}) - \ell_i) - 1, \exp(\ell_{i+1} - \ell_i) - 1\}$, we have the following expression similar to Eqn. (5)

$$\ln z_i = \ln p_i + \ln \int_\Omega f_i(\boldsymbol{\theta}) q(\boldsymbol{\theta}|\ell_i) \mathrm{d}\boldsymbol{\theta} \tag{6}$$

where $\ell_i = \ln l_i$. Suppose we have a random sample $\{\boldsymbol{\Theta}_{i,1}, \boldsymbol{\Theta}_{i,2}, \ldots, \boldsymbol{\Theta}_{i,N}\}$ from the conditional PDF $q(\boldsymbol{\theta}|\ell_i)$, the subarea $z_i$ can be estimated as:

$$\ln \hat{Z}_i = \ln p_i + \ln \left[\frac{1}{N} \sum_{k=1}^{N_i} f_i(\boldsymbol{\Theta}_{i,k})\right] \tag{7}$$

which is unbiased for fixed pairs $(\ell_i, p_i)$. It remains to be issues on how to select $(l_i, p_i)$ and to generate random samples from $q(\boldsymbol{\theta}|\ell_i)$. For these, we proposed to modify the original SuS to adaptively determine $\ell_i$ for a set of exponentially decreasing $p_i$, which will be detailed in the next section.

Note that an alternative way for evaluation of the evidence $z$ of integration can be obtained by continuing Eqn. (4):

$$z = \int_0^{L_{sup}} \int_0^1 \mathbb{1}(p_f(l) > p) \mathrm{d}p \, \mathrm{d}l = \int_0^1 \int_0^{L_{sup}} \mathbb{1}(L_s(p) > l) \, \mathrm{d}l \mathrm{d}p = \int_0^1 L_s(p) \mathrm{d}p \tag{8}$$

where we have used the trick of $p_f(l) = \int_0^1 \mathbb{1}(p_f(l) > p) \mathrm{d}p$ in the first equation and defined the inverse function of $p_f(l)$ as $L_s(p)$ in the second equation. Exchangeability of integration is also assumed to obtain the final expression in terms of $L_s(p)$. Equation (8) indicates the evidence can be equivalently obtained by integrating $L_s(p)$ for $p \in [0,1]$, which explains the main idea behind the NS [12]. Graphically, it represents that the area $z$ can be computed by collecting the horizontal slices enclosed by the dash lines in Figure 1b). Since the length of these slices increases with decreasing $p$ (the typical setting in NS), integrating $L_s(p)$ w.r.t. $p$ may converge slower than integrating $p_f(l)$ w.r.t. $l$, which is the choice of this paper.

Besides the evidence estimation, another crucial task in Bayesian computation is to approximate the posterior distribution. Based on the principle of importance resampling [23], random samples $\{\boldsymbol{\Theta}_{i,1}, \boldsymbol{\Theta}_{i,2}, \ldots, \boldsymbol{\Theta}_{i,N}\}$ generated from $q(\boldsymbol{\theta}|\ell_i)$ can be transformed to posterior samples, via a resampling process according to the weight

$$w_{i,k} = \frac{z^{-1}L(\boldsymbol{\Theta}_{i,k})\pi(\boldsymbol{\Theta}_{i,k})}{p_i^{-1}\pi(\boldsymbol{\Theta}_{i,k})} = \frac{p_i}{z}L(\boldsymbol{\Theta}_{i,k}) \qquad (9)$$

That is, to obtain the equally weighted posterior samples, we just need to accept point $\boldsymbol{\Theta}_{i,k}$ with a probability of $w_{i,k}/\sum_i \sum_k w_{i,k}$. If the task is to evaluate the expectation of a function of the unknown parameter $g(\boldsymbol{\theta})$, it is not necessary to calculate the equally weighted samples, and we can directly utilize the weighted samples such that

$$\mathbb{E}[g(\boldsymbol{\theta})] = \frac{\sum_i \sum_k w_{i,k}\, g(\boldsymbol{\Theta}_{i,k})}{\sum_i \sum_k w_{i,k}} \qquad (10)$$

However, posterior samples based on resampling with weight shown in Eqn. (9) are limited to existing samples, which can be subjected to diversity and ergodicity issue. If more posterior samples are needed, we can use them as seed samples and generate more samples using MCMC.

## 3. Subset simulation for Bayesian analysis

Subset simulation (SuS) is an efficient technique designed to address high-dimensional reliability estimation. It is more resistant to the "curse of dimensionality" as it progresses with thresholds set in the one-dimensional LSF space. In this paper, we modify SuS for Bayesian inference while inheriting its efficiency. In Bayesian analysis, the LSF is related to the likelihood function, and the integration of interest becomes the evidence as described in Eqn. (2). As a result, the original SuS algorithm needs to be adjusted accordingly, and these modifications will be outlined in this section.

### 3.1 Main procedures

To evaluate the evidence based on Eqn. (7), we need to determine pairs of $(\ell_i, p_i)$ and generate random samples from $q(\boldsymbol{\theta}|\ell_i)$ for $i = 0,1,\ldots,M$. For these, the idea of SuS is to set $p_i = p_c^i$ ($p_c$ is called the level probability, a fixed value usually ranging among [0.1, 0.3]) and then to adaptively determine $\ell_i$ based on random samples $\{\boldsymbol{\Theta}_{i-1,1}, \boldsymbol{\Theta}_{i-1,2}, \ldots, \boldsymbol{\Theta}_{i-1,N}\}$, which are generated from $q(\boldsymbol{\theta}|\ell_{i-1})$ [20] via a parallel MCMC scheme. The reason to choose fixed $p_i$ is because its range is known in advance, but the range of likelihood $L(\boldsymbol{\theta})$ can be problem dependent. With this idea, we can iteratively estimate all pairs $(\ell_i, p_i)$ and then estimate the subarea $z_i$ based on Eqn. (7) and finally the overall evidence $z = \sum_{i=0}^{M-1} z_i$. The main procedures of proposed SuS for Bayesian computation are given in Algorithm 1, which consists of initialization, direct MC, parallel MCMC and postprocessing steps.

Algorithm 1: SuS for Bayesian computation
---
1 Initialization
    (1) Given level probability $p_c$, the size of random sample $N$, set number of Markov chains $N_c = Np_c$ and number of samples in each chain $N_s = 1/p_c$;
    (2) Define the change of variable expression $\boldsymbol{\theta} = T(\boldsymbol{U})$;

2. Level 0    % direct MC
    (1) Generate independent normal random samples $\{\boldsymbol{u}_k^{(0)}: k = 1,2,\ldots,N\}$ according to standard normal PDF $\phi_d(\boldsymbol{u})$, and calculate the corresponding log likelihood $y_k^{(0)} = \mathcal{L}\left(T(\boldsymbol{u}_k^{(0)})\right)$.
    (2) Sort $\{y_k^{(0)}: k = 1,2,\ldots,N\}$ in descending order to give the list $\{\ell_k^{(0)}\}$ and set $\ell_1 = \left(\ell_{N_c}^{(0)} + \ell_{N_c+1}^{(0)}\right)/2$. Compute the subarea $\hat{z}_0$ according to Eqn. (7) with $p_0 = 1$ and $l_0 = 0$.

3. For Level $i = 1,2,\ldots, M-1$    % parallel MCMC, see Section 3.2.1
    (1) Collect seeds $\{\boldsymbol{u}_{j0}^{(i)}: j = 1,2,\ldots,N_c\}$ corresponding to log likelihood $\{\ell_j^{(i-1)}, j = 1,\ldots,N_c\}$;
    (2) Adopt the adaptive CS-MH algorithm to generate correlated random samples $\{\boldsymbol{u}_{jt}^{(i)}, j = 1,\ldots,N_c, t = 1,\ldots,N_s\}$, and calculate the corresponding log likelihood $y_{jt}^{(i)} = \mathcal{L}\left(T(\boldsymbol{u}_{jt}^{(i)})\right)$;
    (3) Sort $\{y_{jt}^{(i)}: j = 1,\ldots,N_c, t = 1,\ldots,N_s\}$ in descending order to give the list $\{\ell_k^{(i)}: k = 1,2,\ldots,N_i\}$ and set $\ell_{i+1} = \left(\ell_{N_c}^{(i)} + \ell_{N_c+1}^{(i)}\right)/2$. Compute the subarea $\hat{z}_i$ according to Eqn. (7) with $p_i = p_c^i$.
    (4) If convergence is achieved, STOP; Endif    % see Section 3.2.2
Endfor

4. Postprocessing
    (1) Estimate the evidence $\hat{z} = \sum_{i=0}^{M-1} \hat{z}_i$ and its variance (see Section 3.3);
    (2) regenerate posterior samples $\{\boldsymbol{u}_k: k = 1,2,\ldots,N_{ess}\}$ according to the weight shown in Eqn. (9), and transform back to obtain $\{\boldsymbol{\theta}_k = T(\boldsymbol{u}_k)\}$.    % $N_{ess}$ denotes the effective sample size
---

To initiate the algorithm, two sets of parameters must be established: the level probability $p_c$ and the sample size $N$ in simulation levels $i = 0,1,\ldots M-1$. In the reliability literature, a typical choice is $p_c \in (0.1, 0.3)$ and $N$ is a constant ranging from a few hundreds to over a thousand. We follow this choice in this paper. It is also required that both $N_c = p_c N_i$ and $N_s = p_c^{-1}$ are positive integers. They are respectively equal to the number of chains and the number of samples per chain at simulation levels $i = 1, 2, \ldots, M - 1$ as will be seen shortly.

A change of variable procedure $\boldsymbol{\Theta} = T(\boldsymbol{U})$ is also applied in the initialization. It transforms $\boldsymbol{\Theta} \sim \pi(\boldsymbol{\theta})$ to $\boldsymbol{U} \sim \phi_d(\boldsymbol{u})$, which denotes the standard normal distribution of dimension $d$. Working in the standard normal space is not a burden but provides stability and mathematical convenience. First, it normalizes all parameters in $\boldsymbol{\Theta}$ into the same scale, reducing the possible numerical error. Second, it facilitates the design of efficient MCMC scheme, working for high-dimensional sampling. Various approaches exist for constructing the transformation $T$, e.g., the inverse cumulative distribution function, the Rosenblatt transformation [24] or the marginal transformation based on the Nataf model [25]. In this space, the conditional distribution becomes

$$q(\boldsymbol{u}|\ell_i) = p_i^{-1} \mathbb{1}(\mathcal{L}(\boldsymbol{u}) > \ell_i) \phi_d(\boldsymbol{u}) \tag{11}$$

which is a key target distribution to sample from in SuS.

Following these initializations, the crude MC is then conducted, aiming at estimating $z_0$ for likelihood $l_0 = 0$ and FPF $p_0 = 1$. It involves determining $\ell_1$ satisfying $p_f(l_1) = p_c$, i.e., $\ell_1$ is the $(1 - p_c)$-quantile of the log likelihood $\mathcal{L}$. Given a random sample $\{\boldsymbol{u}_k^{(0)}: k = 1, 2, \ldots, N\}$ generated from $\phi_d(\boldsymbol{u})$, an estimate of $\ell_1$ is then found by sorting the log likelihood values $\{\ell_k^{(0)}: k = 1, \ldots, N\}$ in descending order. Since crude MC is adopted, the generated random sample $\boldsymbol{u}_k^{(0)}$ is independent of each other.

For sampling from $q(\boldsymbol{u}|\ell_i)$ when $i = 1, \ldots, M - 1$, the crude MC becomes infeasible, especially when the dimension $d$ is large. Instead, a parallel MCMC scheme (see Section 3.2.1) is applied, starting from seeds $\{\boldsymbol{u}_{j0}^{(i)}: j = 1, \ldots, N_c\}$. Since these seeds automatically follow the target distribution $q(\boldsymbol{u}|\ell_i)$, no burn-in is needed in the MCMC sampling, saving computational time. Consequently, it enables a strategy of multiple short chains, reducing the correlation between random samples in a single chain. It also potentially improves the ergodicity of generated samples, because switching between different local modes to achieve ergodicity is difficult in MCMC sampling, even for long chains. Once the intended number of samples are generated, their log likelihood values are sorted again to provide an estimate of $\ell_{i+1}$, and thus one can obtain the estimate $\hat{z}_i$. The above parallel MCMC sampling iterates until the termination criteria are satisfied. Please see Section 3.2.2 for a detailed description of the termination condition.

The Bayesian computation, including the evidence estimation and posterior approximation, is arranged as a postprocessing after the adaptive sampling. Since the estimation of evidence is based on random samples, it is helpful to quantify its associated uncertainty. For the posterior resampling from the weighted samples, a critical question is how many 'effective' samples can be generated from the total $MN$ samples. These two issues will be resolved in Section 3.3.

**3.2 Key elements**

The concept behind the SuS algorithm is straightforward, yet its effectiveness can vary depending on how it is implemented. For instance, sampling from a high-dimensional distribution $q(\boldsymbol{u}|\ell_i)$ might pose challenges, and instances of premature convergence could occur. Thus, it is crucial to meticulously design the algorithm.

**3.2.1 Parallel MCMC sampling**

Various MCMC algorithms have been developed for high-dimensional sampling within the SuS framework. Noteworthy among these are component-wise Metropolis-Hastings (MH) [20], conditional sampling MH (CS-MH) [26] and Hamiltonian Monte Carlo sampling [27]. The CS-MH algorithm was proposed according to the fact that the conditional distribution is normally distributed in the multivariate normal distribution. Because of its simplicity and efficiency, the CS-MH algorithm is applied and briefly introduced here for completeness.

To generate a random sample of size $N$ from $q(\boldsymbol{u}|\ell_i)$, the CS-MH algorithm adopts $N_c = Np_c$ parallel MCMC chains, each producing $N_s = 1/p_c$ samples per chain. An adaptive version of CS-MH algorithm is outlined in Algorithm 2. The candidate generation scheme in the inner for-loop is the key for the success of the CS-MH algorithm. Although like the normal proposal in the MH algorithm, the generated sample $\boldsymbol{v}$ automatically follows the standard normal distribution $\phi_n(\boldsymbol{u})$. Since the acceptance of $\boldsymbol{v}$ is contingent solely on whether $\mathcal{L}(\boldsymbol{v}) > \ell_i$, it effectively transforms an $n$-dimensional sampling into one-dimensional problem. To achieve a better balance between ergodicity and efficiency, an adaptive scheme was proposed to optimally choose a scaling parameter $\lambda_{iter}$ such that the average acceptance rate is close to 0.44 based on the one-dimensional Langevin diffusion. Please refer to [26] for more details.

---

**Algorithm 2: Adaptive CS-MH**

---

1. Given the number of chains $N_c$, the length of each chain seeds $N_s$ and the seed $\{\boldsymbol{u}_{j0}: j = 1,2, \ldots, N_c\}$;

Initialize the algorithm by setting $\boldsymbol{\sigma}_0$ to be the sample standard deviation of seeds;

Define initial scaling parameter $\lambda_1 = 0.6$ and the updating frequency $N_a$ ($N_a/N_c \in [0.1, 0.2]$)

2. Permute randomly the seeds $\{\boldsymbol{u}_{j0}: j = 1,2, \ldots, N_c\}$.

3. For $iter = 1,2, \ldots, N_c/N_a$

    (1) Compute the adapted standard deviation $\boldsymbol{\sigma}_{iter} = \min\{\lambda_{iter}\boldsymbol{\sigma}_0, 1\}$ and, subsequently, the correlation coefficient $\boldsymbol{\rho} = \sqrt{1 - \boldsymbol{\sigma}_{iter}^2}$;

    (2) Starting from each seeds $\{\boldsymbol{u}_{[(iter-1)N_a+j]0}: j = 1,2, \ldots, N_a\}$ generate $N_s$ correlated samples $\{\boldsymbol{u}_{[(iter-1)N_a+j]t}: t = 1, \ldots, N_s\}$ from the conditional PDF $q(\boldsymbol{u}|\ell_i)$:

        For $t = 1, \ldots, N_s$

            Generate $\boldsymbol{v} \sim \mathcal{N}\left(\boldsymbol{\rho} \odot \boldsymbol{u}_{[(iter-1)N_a+j](k-1)}, \boldsymbol{I}_d - \mathrm{diag}(\boldsymbol{\rho} \odot \boldsymbol{\rho})\right)$

            % "$\odot$" denotes elementwise product

            If $\mathcal{L}(\boldsymbol{v}) > \ell_i$

                $\boldsymbol{u}_{[(iter-1)N_a+j]k} = \boldsymbol{v}$;

            Else $\boldsymbol{u}_{[(iter-1)N_a+j]k} = \boldsymbol{u}_{[(iter-1)N_a+j](k-1)}$

            Endif

        Endfor

    (3) Evaluate the average acceptance probability $\hat{a}_{iter}$ of the last $N_a$ chains

$$\hat{a}_{iter} = \frac{1}{N_a} \sum_{j=1}^{N_a} \widehat{\mathbb{E}}\left[a(\boldsymbol{u}_{[(iter-1)N_a+j]0})\right]$$

where $\widehat{\mathbb{E}}\left[a(\boldsymbol{u}_{[(iter-1)N_a+j]0})\right]$ is the average accepted number of samples of the chain with seed $\boldsymbol{u}_{[(iter-1)N_a+j]0}$;

    (4) Compute the new scaling parameter

$$\log \lambda_{iter+1} = \log \lambda_{iter} + (\hat{a}_{iter} - 0.44)/\sqrt{iter}$$

Endfor

---

### 3.2.2 Termination condition

A termination condition is necessary to determine the number of iterations $M$, so the remaining unexplored parameter space has negligible contribution to the evidence. In this paper, two heuristic termination conditions are adopted to confirm convergence:

$$\frac{\ell_{i+1} - \ell_i}{\ell_{i+1} + \ell_i} \leq \varepsilon_1; \quad \frac{\hat{Z}_i}{\sum_{ii=1}^{i} \hat{Z}_{ii}} \leq \varepsilon_2 \qquad (12)$$

where $\varepsilon_1$ and $\varepsilon_2$ represent small numbers, which can typically be set to $1 \times 10^{-5}$ $1 \times 10^{-3}$. Eqn. (12) implies that further simulations have a negligible effect on both increasing the likelihood and accumulating the evidence.

If only the first inequality in Eqn. (12) is met, it may lead to false convergence. This scenario is depicted in Figure 2(a), where $p_f(\ell)$ drops sharply, leading to a minimal difference between consecutive thresholds $\ell_i$ and $\ell_{i+1}$ (resembling a plateau in likelihood). If SuS terminates at this point, the estimated evidence overlooks significant contributions from the subsequent subarea $z_{i+1}$ and beyond. Similarly, if only the second inequality in Eqn. (12) is satisfied, as illustrated in Figure 2(b), numerous higher likelihood values remain unexplored (akin to a spike in likelihood). Instances where false convergence satisfies both criteria are uncommon. Hence, employing these dual convergence criteria ensures robustness

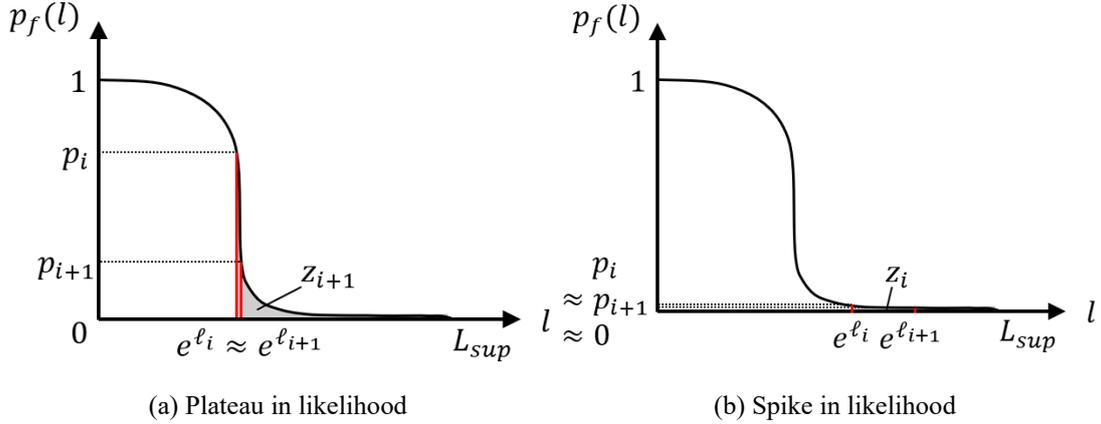

(a) Plateau in likelihood  (b) Spike in likelihood

Figure 2. Two scenarios of false convergence

### 3.3 Performance metrics

Metrics to measure the performance of the SuS algorithm for Bayesian inference are developed in this section, by quantifying the uncertainty associated with the evidence estimator $\hat{Z}$ and the quality of posterior samples $\{\boldsymbol{\Theta}_k\}$. For the former, we propose a strategy to estimate the variance of $\hat{Z}$, while we compute the effective sample size (ESS) $N_{ess}$ for the latter.

### 3.3.1 Estimation variance

From the main procedures of SuS algorithm, we see that the log likelihood threshold $\ell_i$ is, in fact, random, because it is adaptively determined from generated random samples $\{\boldsymbol{\theta}_{jt}^{(i-1)}\}$. However, it is hard to calculate the variance of $\ell_i$ from a single run. Since there is a one-to-one correspondence between $\ell_i$ and $p_i$ via the FPF $p_i = p_f(\ell_i)$, we can equivalently consider $\ell_i$ as a fixed value and estimate $p_i$ in each level. From this perspective and the operation in SuS algorithm, one has the estimator of $p_i$ as $\hat{P}_i = \prod_{ii=0}^{i-1} \hat{P}_c^{(ii)}$, where $\hat{P}_c^{(ii)} = 1/N \sum_{j=1}^{N_c} \sum_{t=1}^{N_s} \mathbb{1}\left[\mathcal{L}\left(\boldsymbol{\theta}_{jt}^{(ii)}\right) > \ell_{ii+1}\right]$ is an unbiased estimator of the level probability $p_c$. One can then rewrite the SuS estimator of $z_i$ as

$$\hat{Z}_i = \hat{H}_i \prod_{ii=-1}^{i-1} \hat{P}_c^{(ii)} \tag{13}$$

where $\hat{H}_i = 1/N \sum_{j=1}^{N_c} \sum_{t=1}^{N_s} f_i\left(\boldsymbol{\theta}_{jt}^{(i)}\right)$ is an unbiased estimator of $h_i = \int_\Omega f_i(\boldsymbol{\theta}) q(\boldsymbol{\theta}|\ell_i) d\boldsymbol{\theta}$. Here, we define $\hat{P}_c^{(-1)} = 1$ to accommodate the case when $i = 0$.

The investigation of the estimation variance of $\hat{Z}$ is involved, because it relates to multiple statistically correlated estimators $\hat{Z}_i$ for $i = 1, 2, \ldots, M-1$, and correlated samples generated from parallel MCMC are used in constructing each estimator. The main results are provided here, and the derivation is postponed into Appendix A. Because of $\hat{Z} = \sum_{i=0}^{M-1} \hat{Z}_i$, one has $\text{VAR}[\hat{Z}] = \sum_{i=0}^{M-1} \sum_{j=0}^{M-1} \text{COV}[\hat{Z}_i, \hat{Z}_j]$, where $\text{COV}[\hat{Z}_i, \hat{Z}_j]$ represents the covariance between estimators $\hat{Z}_i$ and $\hat{Z}_j$. Due to the symmetry of covariance matrix, we only need to evaluate the covariance when $i \leq j$. Assuming random samples generated in distinct levels are statistically independent and the number of samples in each level is large, one can approximate the covariance term by

$$\text{COV}[\hat{Z}_i, \hat{Z}_j] \approx z_i z_j \left\{ \left(\delta_h^{(i)}\right)^2 \mathbb{1}(i=j) + \sum_{ii=0}^{i-1} \left(\delta_p^{(ii)}\right)^2 + \rho_{hp}^{(i)} \delta_h^{(i)} \delta_p^{(i)} \mathbb{1}(i<j) \right\} \tag{14}$$

To calculate this value, $z_i$ and $z_j$ can be approximated using their estimates, $\hat{z}_i$ and $\hat{z}_j$, respectively. Terms $\delta_h^{(i)}$, $\delta_p^{(i)}$ and $\rho_{hp}^{(i)}$ represent the coefficient of variation (c.o.v.) of $\hat{H}_i$, the c.o.v. of $\hat{P}_c^{(i)}$, and the correlation coefficient between them, respectively. Considering the autocorrelation within a Markov chain, they can be estimated as

$$\hat{\delta}_h^{(i)} = \sqrt{\frac{\text{var}[f_i]}{N\hat{h}_i^2}(1+\gamma_h^{(i)})}$$

$$\hat{\delta}_p^{(i)} = \sqrt{\frac{\text{var}[\mathbb{1}_i]}{N\hat{p}_c^{(i)^2}}(1+\gamma_p^{(i)})} \quad (15)$$

$$\hat{\rho}_{hp}^{(i)} = \hat{\rho}_{f\mathbb{1}}^{(i)}\left[1+\gamma_{hp}^{(i)}\right]/\sqrt{\left[1+\gamma_h^{(i)}\right]\left[1+\gamma_p^{(i)}\right]}$$

where $\hat{h}_i$ and $\hat{p}_c^{(i)}$ are the estimates of $\widehat{H}_i$ and $\widehat{P}_c^{(i)}$, respectively; $\text{var}[f_i]$, $\text{var}[\mathbb{1}_i]$ and $\hat{\rho}_{f\mathbb{1}}^{(i)}$ denote the sample variance of functions $f_i(\boldsymbol{\theta})$, sample variance of $\mathbb{1}[\mathcal{L}(\boldsymbol{\theta}) > \ell_{i+1}]$, and their sample correlation coefficient, respectively. Terms $\gamma_h^{(i)}$, $\gamma_p^{(i)}$ and $\gamma_{hp}^{(i)}$ are correlation factors that capture the variance amplification effect due to the autocorrelation and cross-correlation of between the function $f_i(\boldsymbol{\theta})$ and indicator $\mathbb{1}[\mathcal{L}(\boldsymbol{\theta}) > \ell_{i+1}]$ in a Markov chain. The estimations of $\gamma_h^{(i)}$ and $\gamma_p^{(i)}$ have been derived in Ref. [28] and are provided in Appendix A for completeness, where also contains the derivation and estimation of $\gamma_{hp}^{(i)}$.

### 3.3.2 Effective sample size

In addition to quantifying the uncertainty of the evidence estimator $\hat{Z}$, we also need to assess the quality of the posterior sampling, because correlated and weighted samples are directly generated in the SuS estimator. For this, we consider the ESS, which represents the number of independent and equally weighted MC samples that yields the same variance as the SuS estimator in the evidence evaluation [29], i.e.,

$$N_{ess} = MN\frac{\text{VAR}[\tilde{Z}]}{\text{VAR}[\hat{Z}]} \quad (16)$$

Here, $\tilde{Z}$ denotes the evidence estimator based on independent samples from the posterior distribution $p(\boldsymbol{\theta}|\boldsymbol{D})$, and $\hat{Z}$ is the SuS estimator. We further introduce a hypothetical evidence estimator $\check{Z}$, which is based on independent samples but adaptively generated from $q(\boldsymbol{\theta}|\ell_i)$ as in the SuS estimator. Equation (16) can be equivalently written as

$$N_{ess} = MN\frac{\text{VAR}[\tilde{Z}]}{\text{VAR}[\check{Z}]}\frac{\text{VAR}[\check{Z}]}{\text{VAR}[\hat{Z}]} \quad (17)$$

The term $\text{VAR}[\tilde{Z}]/\text{VAR}[\check{Z}]$ can be regarded as a measure of the distance between the mixture distribution $\sum_{i=0}^{M-1} q(\boldsymbol{\theta}|\ell_i)$ and the posterior distribution $p(\boldsymbol{\theta}|\boldsymbol{D})$. The term $\text{VAR}[\check{Z}]/\text{VAR}[\hat{Z}]$ can be interpreted as a reduction factor due to the correlation between samples in the MCMC chains. Following the guidance for the maximum number of equally weighted posterior samples [30], we can approximate Eqn. (17) by

$$N_{ess} \approx \frac{(\sum_i \sum_k w_{i,k})^2}{\sum_i \sum_k w_{i,k}^2} \frac{\text{var}[\check{Z}]}{\text{var}[\hat{Z}]} \quad (18)$$

where $w_{i,k}$ is the posterior weight given by Eqn. (9). The calculation of $\text{var}[\check{Z}]$ is like that of $\text{var}[\hat{Z}]$ discussed in Section 3.3.1 but keeping $\gamma_h^{(i)} = \gamma_p^{(i)} = \gamma_{hp}^{(i)} = 0$ for $i = 1,2,\dots,M-1$.

## 4. Empirical studies

The performance of the proposed SuS algorithm for Bayesian inference is investigated in this section. We first consider three benchmark examples, which works as a platform for comparison with two state-of-art approaches in terms of evidence estimation and posterior sampling. An example on finite element (FE) model updating is then considered, illustrating the performance of SuS for a practical high-dimensional and possibly multi-modal problem.

### 4.1 Three benchmark examples

To validate the effectiveness of SuS for Bayesian inference, we first illustrate its performance in three benchmark problems featured with multi-dimension and multi-modality [13,31,32]. Their two-dimensional scenarios are illustrated in Figure 3.

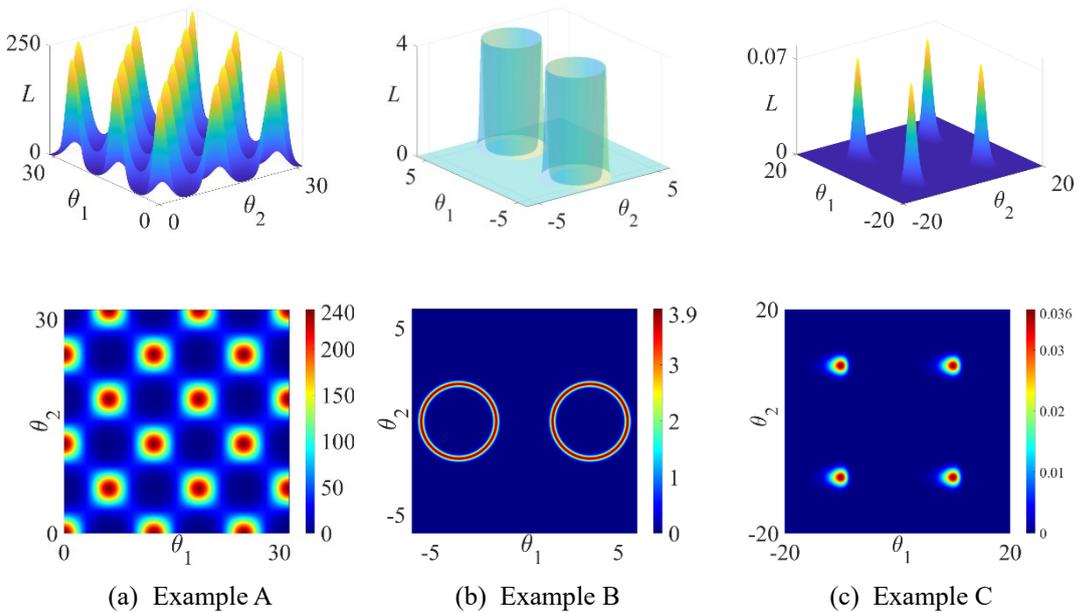

(a) Example A  (b) Example B  (c) Example C

Figure 3. The shape and contour map of likelihood function; Benchmark problems

*Example A: "Eggbox" problem.* Although with only two dimensions, this problem is characterized by an extreme number of modes [33]. The log-likelihood function has the following expression

$$\mathcal{L}(\boldsymbol{\theta}) = \left[2 + \cos\left(\frac{\theta_1}{2}\right)\cos\left(\frac{\theta_2}{2}\right)\right]^5 \tag{19}$$

and a uniform prior $\mathcal{U}(0,10\pi)$ is assumed for both random variables (RVs) $\Theta_1$ and $\Theta_2$.

*Example B: "Normal shells".* This problem graphically represents two well-separated rings in two dimensions [34]. The likelihood function is defined as

$$L(\boldsymbol{\theta}) = \text{circ}(\boldsymbol{\theta}; \boldsymbol{c}_1, r_1, w_1) + \text{circ}(\boldsymbol{\theta}; \boldsymbol{c}_2, r_2, w_2) \tag{20}$$

where

$$\text{circ}(\boldsymbol{\theta}; \boldsymbol{c}, r, w) = \frac{1}{\sqrt{2\pi w^2}} \exp\left[-\frac{(|\boldsymbol{\theta} - \boldsymbol{c}| - r)^2}{2w^2}\right] \tag{21}$$

Here, center vectors $\boldsymbol{c}_1$ and $\boldsymbol{c}_2$ are defined to be $-3.5$ and $3.5$, respectively, in the first dimension, and 0 for the remaining. In addition, $w_1 = w_2 = 0.1$ and $r_1 = r_2 = 2$, and a uniform prior $\mathcal{U}(-6,6)$ is adopted for all RVs $\{\Theta_i\}$.

*Example C: "Normal-LogGamma mixture".* It features four well-separated modes in the first two dimensions [34]. The likelihood function is defined as

$$L(\boldsymbol{\theta}) = \prod_{i=1}^{d} L(\theta_i) \tag{22}$$

where

$$\begin{aligned} L(\theta_1) &= 0.5\text{LogGamma}(\theta_1|10,1,1) + 0.5\text{LogGamma}(\theta_1|-10,1,1) \\ L(\theta_2) &= 0.5\mathcal{N}(\theta_2|10,1) + 0.5\mathcal{N}(\theta_2|-10,1) \end{aligned} \tag{23}$$

for $3 \leq i \leq \frac{d+2}{2}$

$$L(\theta_i) = \text{LogGamma}(\theta_i|10,1,1) \tag{24}$$

and $\frac{d+2}{2} < i \leq d$

$$L(\theta_i) = \mathcal{N}(\theta_i|10,1) \tag{25}$$

Here, "LogGamma" and "$\mathcal{N}$" denote the log Gamma distribution and normal distribution, respectively. A uniform prior $\mathcal{U}(-30,30)$ is assumed for all RVs $\{\Theta_i\}$.

We have chosen another two Bayesian updating algorithms, MULTINEST [35] and aBUS (another SuS-based method) [21], for comparison. First, we investigate the performance of three algorithms in estimating evidence. SuS and aBUS are set with level probability $p_c = 0.1$, and all algorithms are repeated 1000 times. The results are presented in Table 1 in a natural logarithmic scale and, where the MULTINEST results are sourced from Ref. [35].

In Example A, MULTINEST demonstrates superior unbiasedness with comparable computational costs. This is evident as it effectively covers the high likelihood region with a series of ellipticals. However, in the case of two-dimensional normal shells, where the shape of the high likelihood area consists of two rings, MULTINEST results in bias because it is difficult to characterize the high likelihood area using multiple ellipsoidal shapes. On the other hand, SuS and aBUS maintain their unbiased nature. When it comes to uncertainty, the SuS-based algorithm (SuS and aBUS) outperforms the NS-based algorithm (MULTINEST). This difference arises, because the NS algorithm discards one sample at a time, whereas SuS drops 90% of samples in each iteration. As a result, SuS reaches the high likelihood region more quickly, leading to a higher concentration of samples within high likelihood region. Overall, SuS and aBUS demonstrate similar performance due to their shared techniques and surpass MULTINEST in terms of uncertainty and unbiasedness for evidence estimation.

Table 1. Evidence estimation of SuS, aBUS and MULTINEST; Benchmark problems

| Example | Dimension | Analytical | SuS | | | aBUS | | | MULTINEST [35] | | |
|---|---|---|---|---|---|---|---|---|---|---|---|
| | | | Mean | c.o.v. [%] | $N_{cal}$ [$10^3$] | Mean | c.o.v. [%] | $N_{cal}$ [$10^3$] | Mean | c.o.v. [%] | $N_{cal}$ [$10^3$] |
| A | 2 | 235.86 | 235.81 | 0.13 | 19.0 | 235.83 | 0.14 | 19.1 | 235.85 | 0.33 | 20.0 |
| | 2 | -1.75 | -1.75 | 4.00 | 4.40 | -1.75 | 5.14 | 4.47 | -1.61 | 5.59 | 4.58 |
| | 5 | -5.67 | -5.67 | 2.47 | 8.80 | -5.67 | 2.82 | 8.82 | -5.42 | 2.77 | 8.92 |
| B | 10 | -14.59 | -14.58 | 0.96 | 72.0 | -14.57 | 0.89 | 72.3 | -14.55 | 1.58 | 73.3 |
| | 20 | -36.09 | -35.96 | 0.67 | 213 | -35.95 | 0.64 | 202 | -35.90 | 0.97 | 219 |
| | 30 | -60.13 | -59.85 | 0.47 | 548 | -59.87 | 0.48 | 544 | -59.72 | 0.59 | 549 |
| C | 20 | -81.89 | -81.86 | 1.01 | 2490 | -81.86 | 1.00 | 2200 | -78.84 | 0.51 | 2780 |

In addition to the evidence estimation, the quality of posterior samples generated in aBUS and SuS is compared in terms of ESS $N_{ess}$. The derivation of $N_{ess}$ in aBUS can be found in Appendix B. With level probability $p_c = 0.1$ and a sample size of $N = 1000$ in each iteration, 1000 independent runs are conducted. To ensure a fair comparison, Table 2 displays the results as $N_{ess}/N_{cal}$, where $N_{cal}$ represents the number of likelihood function calls. This ratio represents the equivalent number of independent posterior samples generated per likelihood function evaluation. For dimensions of 2-10, performance of SuS and aBUS are similar. However, for dimensions 20-30 of Examples B and C, SuS consistently achieves higher $N_{ess}/N_{cal}$ values compared to aBUS. It may be attributed to the fact that SuS is able to obtain posterior samples from every iteration, whereas aBUS obtains posterior samples only from the final iteration. As the number of iterations increases, the advantage of SuS in terms of sample quality becomes more apparent. Moreover, the c.o.v. of the estimated ratio $N_{ess}/N_{cal}$ is listed in Table 2. Although both algorithms yield c.o.v. less than 20% SuS gives lower values for all considered examples.

Table 2. $N_{ess}/N_{cal}$ ratio of SuS and aBUS; Benchmark problems

| Example | Dimension | SuS | | aBUS | |
|---|---|---|---|---|---|
| | | Mean [%] | c.o.v. [%] | Mean [%] | c.o.v. [%] |
| A | 2 | 4.00 | 12.62 | 5.83 | 16.12 |
| B | 2 | 24.96 | 4.33 | 23.88 | 15.39 |
| | 5 | 13.25 | 3.19 | 10.89 | 9.26 |
| | 10 | 4.35 | 6.32 | 5.04 | 7.14 |
| | 20 | 1.91 | 9.19 | 1.80 | 16.67 |
| | 30 | 1.36 | 2.94 | 1.02 | 1.98 |
| C | 2 | 12.44 | 12.26 | 14.88 | 17.88 |
| | 5 | 5.97 | 12.56 | 5.59 | 15.75 |
| | 10 | 3.39 | 11.80 | 2.56 | 16.41 |
| | 20 | 1.83 | 6.01 | 1.16 | 14.66 |
| | 30 | 1.68 | 9.52 | 0.82 | 17.07 |

The performance of SuS is also illustrated in the histogram plot of generated samples in Figure 4, representing the results of Example C with dimension 20. The lower left half of the figure displays the simulation results of SuS, while the upper right half shows results of aBUS. For graphical illustration, parameters of dimensions 1-3, 11, 12, and 20 are plotted in Figure 4. The gray and green bars represent the marginal distributions approximated with SuS and aBUS posterior samples, respectively, displayed along the diagonal line. Overlaid on these bars are red curves that represent the analytical marginal PDF curves. In the non-diagonal positions of the figure, heat maps are used to represent the density values for any two dimensions. In addition, the contours of the analytic joint PDF are plotted in terms of red curves. In the histogram of posterior simulation in aBUS, it is seen that one specific bar could occasionally appear abnormally higher than the remaining. This is because repeated samples are produced in the final iteration of aBUS. In the proposed SuS approach, those abnormally high bars have been reduced, because the resampling in SuS efficiently goes across various iterations, enhancing the diversity of posterior samples.

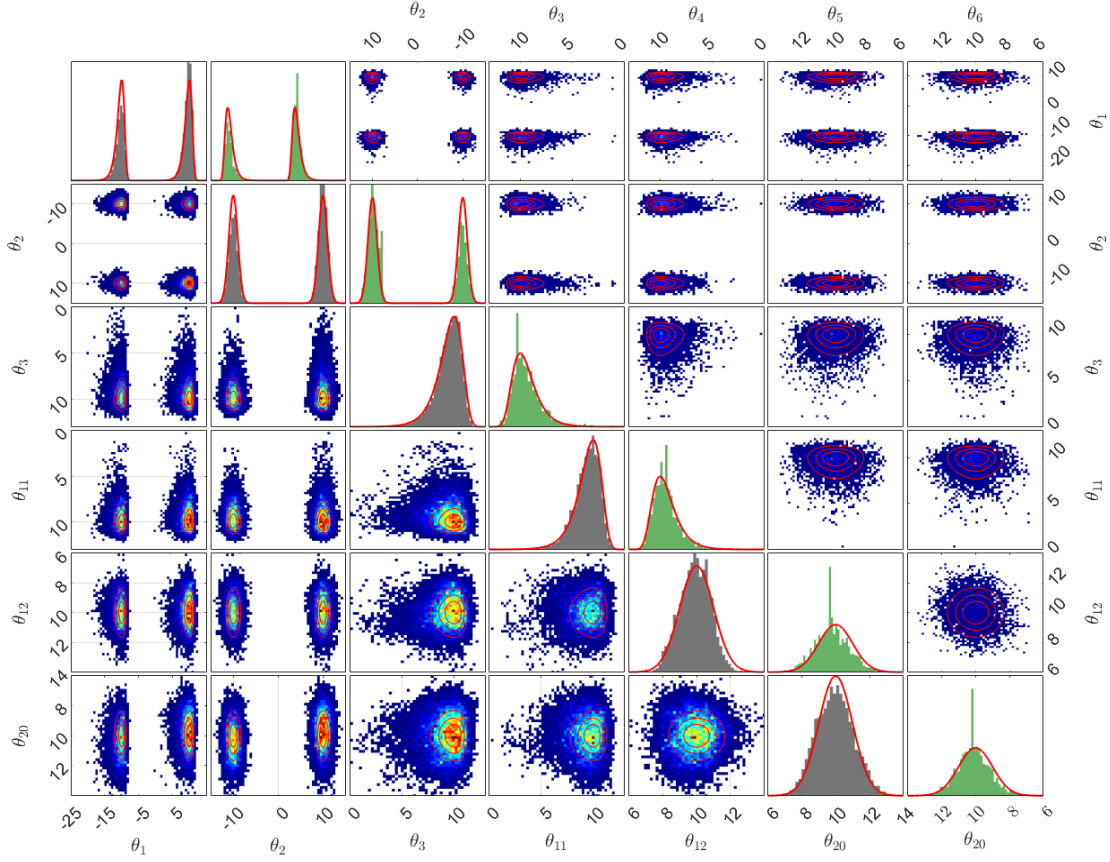

Figure 4. Posterior samples of SuS and aBUS; Dimension 20 in Example C

## 4.2 Finite element model updating

In the application of SuS for FE model updating, we consider a 10-story shear-type building model, as shown in Figure 5, and update it using synthetic ambient vibration data. Initially, the story stiffness and mass are set to be $k_0 = 2 \times 10^6$ kN/m and 1000 tons. Our objective is to modify the stiffness matrix of the FE model to match the simulated response with the ambient vibration data. A factor $\boldsymbol{\alpha} = \alpha_{1:N_d}$ ($N_d = 10$ indicates the degrees of freedom) is introduced as the parameter to be adjusted, resulting in an updated story stiffness profile of $k_0 \alpha_{1:N_d}$.

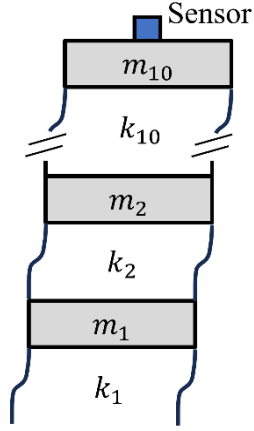

Figure 5. 10-story shear model

Without loss of generality, consider that the structural acceleration response $\{X_t \in \mathbb{C}^{N_c}, t = 1,2,\ldots,T_d\}$ under ambient excitation can be measured with a sampling frequency of $f_s$ (Hz) and a duration of $T_d/f_s$ (sec). Here, $N_c$ denotes the number of measurement channels. They are first divided into $M$ non-overlapping segments and assumed to be independent. Let $\{X_j^r \in \mathbb{C}^{N_c}, j = 1,2,\ldots,T_d/M\}$ denote the $r$-th segment with the following discrete Fourier transform

$$F_k^r = \sqrt{\frac{2M\Delta t}{T_d}} \sum_{j=1}^{T_d/M} X_j^r \exp[-i2\pi M(j-1)(k-1)/T_d], \text{ for } k = 1,2,\ldots,T_d/M \quad (26)$$

corresponding to the frequency $f_k = f_s M(k-1)/T_d$ (Hz), where "$i$" denotes the unit imaginary number. The sample power spectral density (PSD), defined as $\widehat{E}_k = \frac{1}{M}\sum_{r=1}^{M} F_k^r F_k^{r*}$ ("$\cdot^*$" denotes the complex transpose), then asymptotically follow a complex Wishart distribution of dimension $N_c$ and with $M$ degrees of freedom and the mean matrix [35]

$$E_k(\theta) = \Phi h_k S h_k^* \Phi^T + S_e \quad (27)$$

where $h_k$ denotes a diagonal matrix consisting of frequency response functions $h_{ik} = [(1 - \beta_{ik}^2) - i(2\zeta_i \beta_{ik})]^{-1}$ ($\beta_{ik} = f_i/f_k$). Here, $\Phi = [\phi_1, \phi_2, \ldots, \phi_{N_m}] \in \mathbb{R}^{N_c \times N_m}$ represents the partial mode shape matrix confined to the location of measurement channels for a total of $N_m$ modes. Symbols $f_i$ and $\zeta_i$ are the $i$-th natural frequency and damping ratio of the considered structure. Note that mode shape $\Phi$ and natural frequency $f = [f_1, f_2, \ldots, f_{N_m}]$ are functions of the unknown parameter $\alpha$ in terms of the generalized eigenvalue decomposition. The modal force PSD matrix (per unit mass), $S = \text{diag}(S_1, S_2, \ldots, S_{N_m})$, is modeled as a diagonal matrix (i.e., no closely-spaced modes are considered). The error PSD $S_e = \text{diag}(S_{e,1}, S_{e,2}, \ldots, S_{e,N_c})$ denotes the noise level in each channel, accounting for both model inaccuracies and measurement errors. In summary, parameters $\boldsymbol{\theta}$ to be updated include $\{\boldsymbol{\alpha}, \boldsymbol{\zeta}, \boldsymbol{S}, \boldsymbol{S}_e\}$, with a total dimension of $N_d + 2N_m + N_c$. Since $\widehat{\boldsymbol{E}}_k$'s at different frequencies is statistically independent when $T_d/M$ is large enough, one has the following negative log-likelihood function (NLLF) [35]

$$\mathcal{L}(\boldsymbol{\theta}) = c + \sum_{k=1}^{N_f} \ln \det[\boldsymbol{E}_k(\boldsymbol{\theta})] + \sum_{k=1}^{N_f} \text{tr}\big[\boldsymbol{E}_k(\boldsymbol{\theta})^{-1} \widehat{\boldsymbol{E}}_k\big] \tag{28}$$

where $c$ is a constant independent of parameters $\boldsymbol{\theta}$, and $N_f$ denotes the total number of frequency points.

In synthetic data generation, we randomly select the stiffness parameter $\widetilde{\boldsymbol{\alpha}} = [0.71, 0.84, 0.57, 0.78, 0.84, 0.80, 0.93, 0.89, 0.76, 0.76]^T$ to simulate the structural damage. In addition, we set true values of modal force PSD $\widetilde{S}_1 = \cdots = \widetilde{S}_{N_m} = 10^{-10} \text{g}^2/\text{Hz}$ and error PSD $\widetilde{S}_{e,1} = \cdots = \widetilde{S}_{e,N_c} = 10^{-10} \text{g}^2/\text{Hz}$. Gaussian white noise with the above PSDs is generated randomly, and modal superposition method is then adopted to compute the structural acceleration responses with the damping ratio $\widetilde{\zeta}_1 = \cdots = \widetilde{\zeta}_{N_m} = 0.01$. It is important to note that certain preconditions must be met for the NLLF in Eqn. (28) to be valid and appropriate. The data should be generated with a sufficient time length and averaging segments, as referenced in [36]. In this example, the dataset $\boldsymbol{D} = \{\widehat{\boldsymbol{E}}_{1:N_f}\}$ is stored with a frequency resolution of 0.1Hz, as shown in Figure 6 in both PSD and singular value (SV) spectrum (i.e., eigenvalues of PSD matrix), where peaks indicate the location of each mode. Ten modes can be observed from the PSD and SV spectra.

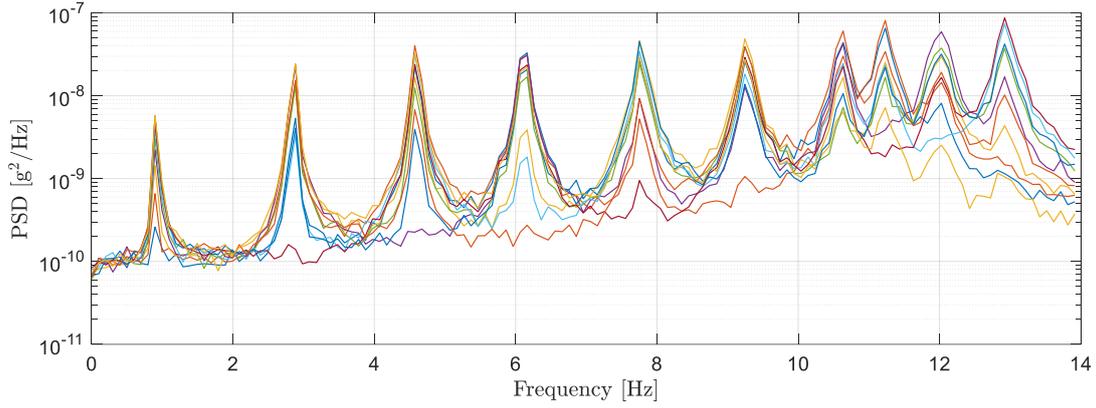

(a) PSD spectrum

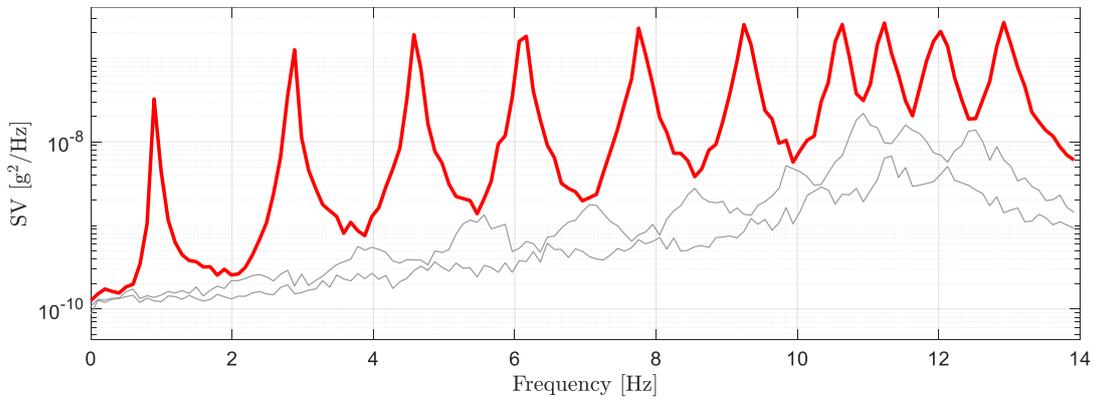

(b) SV spectrum

Figure 6. Frequency-domain representation of synthetic data; FE model updating

Table 3. Test cases; FE model updating

| No. | Measured stories | Adopted modes | Dimension | $N_f$ |
|---|---|---|---|---|
| Case 1 | 9, 10 | 1-5 | 22 | 80 |
| Case 2 | 9, 10 | 1-10 | 32 | 140 |
| Case 3 | 4, 7, 10 | 1-5 | 23 | 80 |
| Case 4 | 4, 7, 10 | 1-10 | 33 | 140 |
| Case 5 | 1 - 10 | 1-5 | 30 | 80 |
| Case 6 | 1 - 10 | 1-10 | 40 | 140 |

For a thorough investigation, six cases are considered (Table 3), featuring various sensor configurations and numbers of modes. The number and location of sensors as well as the number of modes directly determine the identifiability of the problem. It is normal that there might be multiple modes when the configuration of sensors is not appropriate. The total dimensions of parameters are displayed in the fourth column. Their difference lies in the dependence of $\zeta$ and $S$ on the number of modes and the dependence of $S_e$ on the number of measurement channels. The number of modes used for model updating is controlled by the adopted number of frequency points $N_f$.

Table 4. Prior distribution of random variables; FE model updating

| Variables | Distribution | Lower bound | Upper bound |
|---|---|---|---|
| $\alpha_1, \ldots, \alpha_{N_d}(1)$ | Uniform | 0.5 | 1.0 |
| $S_{e,1}, \ldots, S_{e,N_c}(g^2/\text{Hz})$ | Uniform | 0 | $10^{-8}$ |
| $\zeta_1, \ldots, \zeta_{N_m}(1)$ | Uniform | 0 | 0.1 |
| $S_1, \ldots, S_{N_m}(g^2/\text{Hz})$ | Uniform | 0 | $10^{-8}$ |

The developed SuS algorithm is then adopted for FE model updating, with uniform distributions of different parameters (Table 4) used as the prior distribution for Bayesian inference. In the SuS algorithm, we set the level probability to be $p_c = 0.1$ for all cases. Figure 7 displays random samples from the posterior distribution of stiffness parameter $\boldsymbol{\alpha}$, with the red dashed line indicating the predefined "true" values. The displayed coordinate ranges are between 0.5 and 1, aligning with the prior distribution of $\boldsymbol{\alpha}$.

It is evident that inadequate sensor arrangement of Case 1 leads to multiple modes, or even bias in the posterior distribution of stiffness parameter $\boldsymbol{\alpha}$. However, we are still able to precisely identify $\alpha_{10}$, because two sensors measuring Floors 9 and 10 surround the element with parameter $\alpha_{10}$. This suggests that placing sensors near the targeted element is beneficial, although the overall parameters are not globally identifiable. In Case 2, where more structural modes are used for inference, the identification uncertainty decreases because of more data, but it does not improve the biasness. Proper placement of sensors is more critical for an unbiased estimation, as illustrated in Case 3. Although it still shows a large uncertainty, the high probability regions contain the true values of the structural stiffness parameters. When more structural modes are included in Case 4, the identification uncertainty further decreases, yielding a satisfiable performance for engineering application. Cases 5 and 6 illustrate some ideal cases, where we have a large enough number of sensors and modes for inference. In these cases, we can accurately and precisely identify the structural stiffness parameters, illustrating the limit of identification precision. Comparing the results for all cases, it shows adding sensors are more helpful than adding data (in structural modes), and the results may be unreliable if only a few sensors are used but targeting for many parameters.

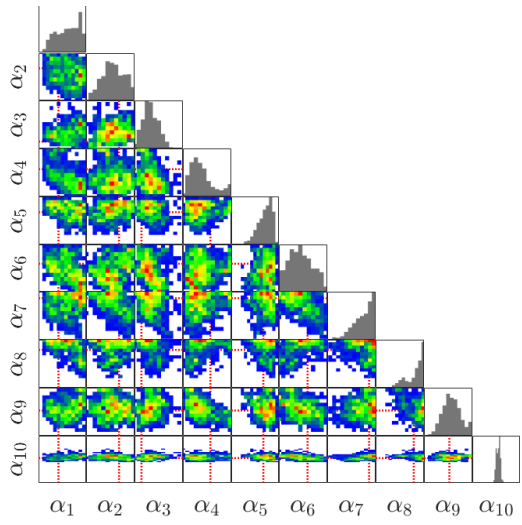

(a) Case 1

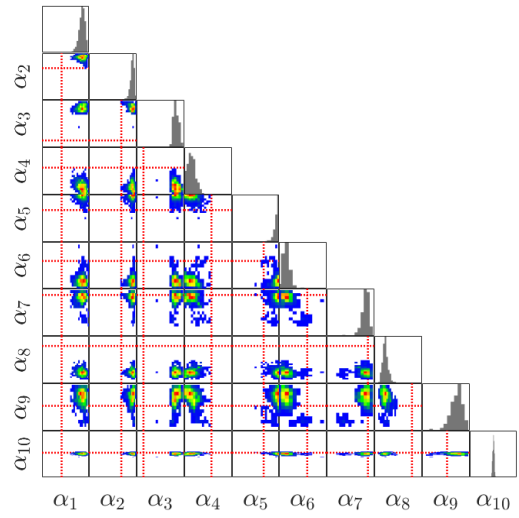

(b) Case 2

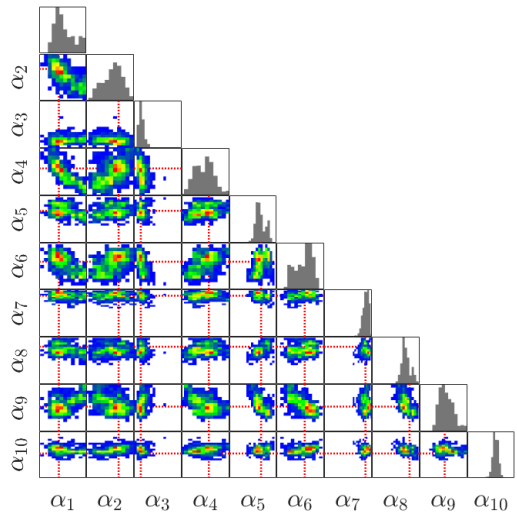

(c) Case 3

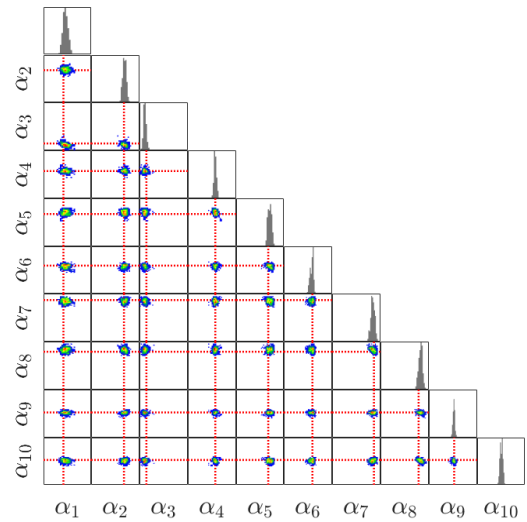

(d) Case 4

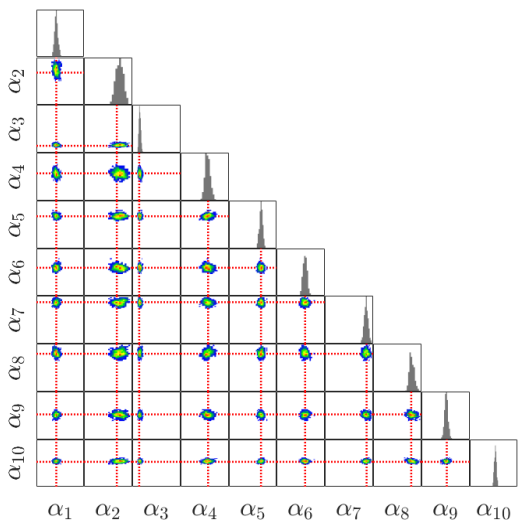

(e) Case 5

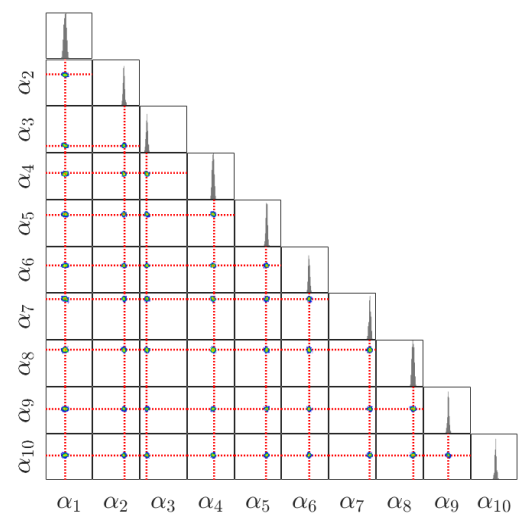

(f) Case 6

Figure 7. Posterior samples of $\boldsymbol{\alpha}$; FE model updating

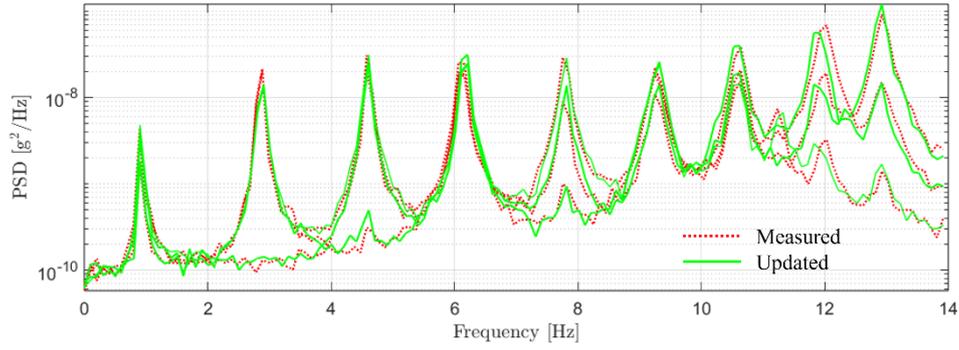

(a) PSD spectrum

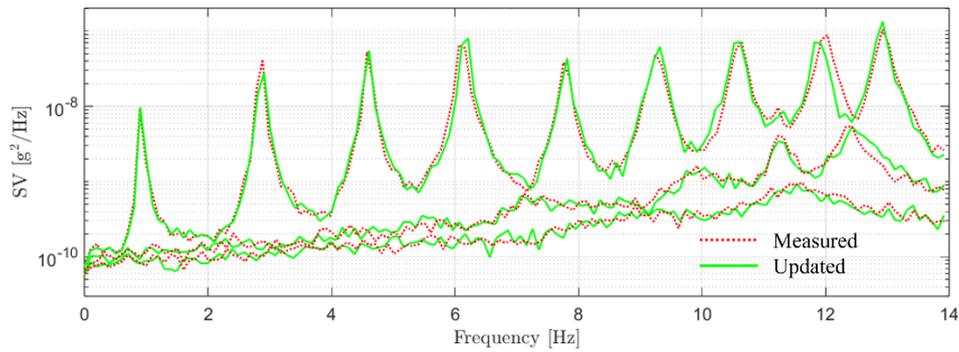

(b) SV spectrum

Figure 8. Frequency-domain representation of dynamics responses for updated model; Case 4

Since Case 4 corresponds to a more practical situation, we provide more analysis results below. First, the measured and updated PSD and SV spectra of the shear-type building model are plotted in Figure 8, where red dashed lines represent measured values, and green solid lines represent the updated spectra correspond to the mean value of the identified stiffness parameters. The updated response matches closely with the measured data. In terms of numbers, a comparison between the posterior mean and the true value of modal parameters is listed in Table 5. It shows the natural frequencies of the updated model are close to the true values, and the associated uncertainty is small. However, there are biases in the estimated damping ratios $\zeta$ and modal force PSDs $S$, which might be due to the modeling error, e.g., the neglect of influence from neighboring modes. The large uncertainty in the estimated values of $\zeta$ and $S$ can also partially explain the estimation bias, because it is easy to see that all true values lie within the 90% credible interval of the Bayesian estimation.

Table 5. Bayesian updating results in Case 4

| Mode | Natural frequency | | | Damping ratio | | | Modal force PSD | | |
|---|---|---|---|---|---|---|---|---|---|
| | True [Hz] | Mean [Hz] | c.o.v. [%] | True [%] | Mean [%] | c.o.v. [%] | True [$(\mu g)^2$/Hz] | Mean [$(\mu g)^2$/Hz] | c.o.v. [%] |
| 1 | 0.920 | 0.920 | 0.342 | 1.000 | 1.964 | 67.46 | 100.0 | 117.6 | 30.68 |
| 2 | 2.848 | 2.857 | 0.381 | 1.000 | 1.044 | 52.99 | 100.0 | 103.6 | 28.60 |
| 3 | 4.594 | 4.591 | 0.501 | 1.000 | 1.035 | 57.74 | 100.0 | 122.1 | 21.13 |
| 4 | 6.114 | 6.118 | 0.381 | 1.000 | 1.266 | 46.34 | 100.0 | 102.9 | 23.03 |
| 5 | 7.784 | 7.762 | 0.423 | 1.000 | 1.463 | 38.96 | 100.0 | 131.9 | 31.25 |
| 6 | 9.268 | 9.252 | 0.409 | 1.000 | 1.152 | 44.35 | 100.0 | 120.6 | 21.68 |
| 7 | 10.609 | 10.608 | 0.385 | 1.000 | 1.592 | 25.64 | 100.0 | 124.5 | 26.27 |
| 8 | 11.218 | 11.287 | 0.688 | 1.000 | 4.451 | 65.56 | 100.0 | 51.3 | 76.70 |
| 9 | 11.993 | 11.973 | 0.361 | 1.000 | 1.493 | 31.19 | 100.0 | 145.3 | 15.60 |
| 10 | 12.941 | 12.962 | 0.365 | 1.000 | 1.334 | 21.83 | 100.0 | 131.6 | 17.19 |

## 5. Conclusions

In this paper, we develop an efficient algorithm for Bayesian inference of high-dimensional and multi-modal problems by interpreting the evidence estimation as a sequential reliability estimation problem. The subset simulation (SuS) algorithm is then adopted to estimate the evidence, and posterior samples are generated following the principle of importance resampling. The uncertainty associated with the estimated evidence is quantified by estimating the variance. The effective sample size is also computed to measure the performance of posterior sampling. Three benchmark examples and one FE model updating problem are considered to illustrate the performance of the proposed SuS algorithm, by comparing it with two state-of-art algorithms.

The proposed algorithm exhibits comparable or even better performance in evidence estimation and posterior sampling, compared to the aBUS and MULTINEST algorithms, especially when the dimension of unknown parameters is high. In the application of the proposed algorithm for FE model updating, satisfactory performance can be obtained when the configuration (number and location) of sensory system is proper. The sensor place plays a more critical role than the number of structural modes used for identification. For a specific sensor configuration, some parameters near the sensor locations may still be identifiable, although the overall problem is not globally identifiable.

The setting of parameters in the proposed SuS algorithm follows the optimal values determined for reliability estimation, and thus it may not be optimal for the case of Bayesian inference. Since the formulae for variance of estimated evidence and the effective sample size have been derived, they provide a foundation for the further improvement of the proposed SuS algorithm for Bayesian inference.

## CRediT authorship contribution statement

**Zihan Liao**: Writing - review & editing, Writing - original draft, Visualization, Validation, Software, Methodology, Investigation, Formal analysis, Data curation, Conceptualization.

**Binbin Li**: Writing - review & editing, Writing - original draft, Visualization, Validation, Supervision, Software, Resources, Project administration, Methodology, Investigation, Funding acquisition, Formal analysis, Data curation, Conceptualization.

**Hua-Ping Wan**: Writing - review & editing, Supervision, Resources, Project administration, Methodology, Investigation, Formal analysis, Conceptualization.

## Declaration of competing interest

The authors declare that they have no known competing financial interests or personal relationships that could have appeared to influence the work reported in this paper.

## Data availability

The data and code will be made available once the paper is accepted for publication.

## Acknowledgement

This work was led by Principal Supervisor Binbin Li, and it was partially supported by the National Natural Science Foundation of China (U23A20662 and 52422804) and the Center for Infrastructure Resilience in Cities as Livable Environments (CIRCLE) through the Zhejiang University-University of Illinois Urbana-Champaign Joint Research Center Project No. DREMES202001, funded by Zhejiang University. Any opinions, findings, conclusions, or recommendations expressed in this material are those of the authors and do not reflect the views of the funders.

# Appendix A. Estimation variance of $\widehat{Z}$

Before jumping into the formula in calculating the overall variance, we first analyze the source of uncertainties in the estimator $\hat{Z}_i$. From Eqn. (13), we can see that the estimator $\hat{Z}_i$ can be decomposed into multiple estimators $\hat{P}_c^{(ii)}$ and $\widehat{H}_i$. With their unbiasedness, we model them as $\hat{P}_c^{(ii)} = p_c(1 + E_{ii})$ and $\widehat{H}_i = h_i(1 + \Omega_i)$, where $E_{ii}$ and $\Omega_i$ denote two zero-mean RVs. It is easy to see that the variance of $E_{ii}$ is equal to the squared c.o.v. of $\hat{P}_c^{(ii)}$, i.e., $\text{VAR}[E_{ii}] = \left(\delta_p^{(ii)}\right)^2$. Similarly, we have $\text{VAR}[\Omega_i] = \left(\delta_h^{(i)}\right)^2$, where $\delta_r^{(i)}$ denotes the c.o.v. of $\widehat{H}_i$. Here, we assume that the size of random sample $N$ is large enough to yield small c.o.v.s for both $\hat{P}_c^{(ii)}$ and $\widehat{H}_i$, so that $E_{ii}$ and $\Omega_i$ can be regarded as small variations from zero. Ignoring high order terms of $E_{ii}$ and $\Omega_i$, one can obtain the following linear approximation

$$\hat{Z}_i \approx p_c^i h_i \left(1 + \Omega_i + \sum_{ii=0}^{i-1} E_{ii}\right) \tag{A.1}$$

Because $\text{VAR}[\hat{Z}] = \sum_{i=0}^{M-1} \sum_{j=0}^{M-1} \text{COV}[\hat{Z}_i, \hat{Z}_j]$, the key problem lies in how to evaluate the covariance $\text{COV}[\hat{Z}_i, \hat{Z}_j] = \mathbb{E}[\hat{Z}_i \hat{Z}_j] - z_i z_j$. Assuming random samples generated in distinct levels are statistically independent, and substituting the linear approximation of $\hat{Z}_i$ and $\hat{Z}_j$, one can show that

$$\mathbb{E}[\hat{Z}_i \hat{Z}_j] \approx z_i z_j \left\{1 + \mathbb{E}[\Omega_i^2]\mathbb{1}(i=j) + \sum_{ii=0}^{i-1} \mathbb{E}[E_{ii}^2] + \mathbb{E}[\Omega_i E_i]\mathbb{1}(i<j)\right\} \tag{A.2}$$

Substituting $\text{VAR}[E_{ii}] = \left(\delta_p^{(ii)}\right)^2$ and $\text{VAR}[\Omega_i] = \left(\delta_h^{(i)}\right)^2$, one can obtain Eqn. (14).

For Eqn. (15), the first two equations were previously derived in [28], and the derivation of the third equation is provided below. According to the definition, the correlation coefficient between the estimators $\hat{P}_c^{(i)}$ and $\widehat{H}_i$ can be expressed as $\rho_{hp}^{(i)} = \text{COV}\left[\widehat{H}_i, \hat{P}_c^{(i)}\right] / \sqrt{\text{VAR}[\widehat{H}_i]\text{VAR}\left[\hat{P}_c^{(i)}\right]}$. The covariance $\text{COV}\left[\widehat{H}_i, \hat{P}_c^{(i)}\right]$ is expanded as

$$\begin{aligned}\text{COV}\left[\widehat{H}_i, \hat{P}_c^{(i)}\right] &= \frac{1}{N^2} \mathbb{E}\left[\left\{\sum_{j=1}^{N_c}\sum_{t=1}^{N_s} f_i^{(j,t)}\right\}\left\{\sum_{j=1}^{N_c}\sum_{t=1}^{N_s} \mathbb{1}_i^{(j,t)}\right\}\right] - h_i p_c \\ &= \frac{1}{N^2} \sum_{j_1=1}^{N_c}\sum_{t_1=1}^{N_s}\sum_{j_2=1}^{N_c}\sum_{t_2=1}^{N_s} \text{COV}\left[f_i^{(j_1,t_1)}, \mathbb{1}_i^{(j_2,t_2)}\right]\end{aligned} \tag{A.3}$$

where we have used the simplified notations $f_i^{(j,t)} = f_i\left(\boldsymbol{\Theta}_{jt}^{(i)}\right)$ and $\mathbb{1}_i^{(j,t)} = \mathbb{1}\left[\mathcal{L}\left(\boldsymbol{\Theta}_{jt}^{(i)}\right) > \ell_{i+1}\right]$. Assuming statistical independence between samples generated from different Markov chains, i.e., $\text{COV}\left[f_i^{(j_1,t_1)}, \mathbb{1}_i^{(j_2,t_2)}\right] = 0$ for any $j_1 \neq j_2$, and identical autocorrelation within each Markov chain, i.e., $\text{COV}\left[f_i^{(j_1,t_1)}, \mathbb{1}_i^{(j_1,t_2)}\right] = \text{COV}\left[f_i^{(j_2,t_1)}, \mathbb{1}_i^{(j_2,t_2)}\right] = \text{COV}[f_i^{(t_1)}, \mathbb{1}_i^{(t_2)}]$, the above equation can be simplified as

$$\text{COV}\left[\widehat{H}_i, \widehat{P}_c^{(i)}\right] = \frac{N_c}{N^2} \sum_{t_1=1}^{N_s} \sum_{t_2=1}^{N_s} \text{COV}\left[f_i^{(t_1)}, \mathbb{1}_i^{(t_2)}\right] \tag{A.4}$$

which reduces the problem to calculating the covariance $\text{COV}\left[f_i^{(t_1)}, \mathbb{1}_i^{(t_2)}\right]$.

Taking advantage of the stationarity of MCMC, the variances of samples along the chain does not change, i.e., $\text{VAR}\left[f_i^{(t_1)}\right] = \text{VAR}[f_i]$ and $\text{VAR}\left[\mathbb{1}_i^{(t_2)}\right] = \text{VAR}[\mathbb{1}_i]$. Therefore, we can obtain the following expression for the correlation coefficient between $\widehat{P}_c^{(i)}$ and $\widehat{H}_i$

$$\begin{aligned}\rho_{hp}^{(i)} &= N_c \sum_{t_1=1}^{N_s} \sum_{t_2=1}^{N_s} \frac{\text{COV}\left[f_i^{(t_1)}, \mathbb{1}_i^{(t_2)}\right]}{\sqrt{\text{VAR}[f_i]\left(1 + \gamma_h^{(i)}\right) \text{VAR}[\mathbb{1}_m]\left(1 + \gamma_p^{(i)}\right)}} \\ &= \frac{1}{\sqrt{\left(1 + \gamma_h^{(i)}\right)\left(1 + \gamma_p^{(i)}\right)}} \frac{1}{N_s} \sum_{t_1=1}^{N_s} \sum_{t_2=1}^{N_s} \rho\left[f_i^{(t_1)}, \mathbb{1}_i^{(t_2)}\right]\end{aligned} \tag{A.5}$$

where $\rho\left[f_i^{(t_1)}, \mathbb{1}_i^{(t_2)}\right]$ is the correlation between $f_i^{(t_1)}$ and $\mathbb{1}_i^{(t_2)}$. The correlation factors $\gamma_h^{(i)}$ and $\gamma_p^{(i)}$ capture the variance amplification effect resulting from the autocorrelation of function $f_i^{(t)}$ and indicator $\mathbb{1}_i^{(t)}$ within Markov chains. To quantify these correlation factors, an approximation formula has been proposed in Ref. [28]: $\gamma_h^{(i)} \approx g\left(\rho_h^{(i)}\right)$, $\gamma_p^{(i)} = g\left(\rho_p^{(i)}\right)$ with the function $g(\cdot)$ expressed as

$$g(\rho) = 2\rho\{1 - \rho - [1 - \rho^{N_s}]/N_s\}/[1-\rho]^2 \tag{A.6}$$

and $\rho_h^{(i)}$ and $\rho_p^{(i)}$ being the lag 1 correlation of $f_i^{(t)}$ and $\mathbb{1}_i^{(t)}$, respectively. They can be estimated from the generated random samples as

$$\widehat{\rho}_h^{(i)} = \frac{1}{N_s - 1} \sum_{t=1}^{N_s-1} \frac{\frac{1}{N_c}\sum_{j=1}^{N_c} f_i^{(j,t)} f_i^{(j,t+1)} - \widehat{h}_i^2}{\frac{1}{N_c}\sum_{j=1}^{N_c} f_i^{(j,t)^2} - \widehat{h}_i^2} \tag{A.7}$$

$$\hat{\rho}_{\mathbb{1}}^{(i)} = \frac{1}{N_s - 1} \sum_{t=1}^{N_s-1} \frac{\frac{1}{N_c}\sum_{j=1}^{N_c} \mathbb{1}_i^{(j,t)} \mathbb{1}_i^{(j,t+1)} - \hat{p}_c^{(i)2}}{\frac{1}{N_c}\sum_{j=1}^{N_c} \mathbb{1}_i^{(j,t)2} - \hat{p}_c^{(i)2}}$$

For the correlation $\rho\left[f_i^{(t_1)}, \mathbb{1}_i^{(t_2)}\right]$, we model it as the multiplication of the cross-correlation (with no lag) between functions $f_i^{(t)}$ and $\mathbb{1}_i^{(t)}$ (denoted as $\rho_{f\mathbb{1}}^{(i)}$) and a part accounting the lag effect, i.e.,

$$\rho\left[f_i^{(t_1)}, \mathbb{1}_i^{(t_2)}\right] = \rho_{f\mathbb{1}}^{(i)} \left[\rho_{f\mathbb{1},c}^{(i)}\right]^{|t_1-t_2|} \tag{A. 8}$$

where we have assumed an exponential decay of correlation as the lag $|t_1 - t_2|$ increases [28]. Substituting Eqn. (A. 8) into Eqn. (A. 5) yields

$$\rho_{hp}^{(i)} = \rho_{f\mathbb{1}}^{(i)} \left[1 + \gamma_{hp}^{(i)}\right] / \sqrt{\left[1 + \gamma_h^{(i)}\right]\left[1 + \gamma_p^{(i)}\right]} \tag{A. 9}$$

where $\gamma_{hp}^{(i)} = g\left(\rho_{f\mathbb{1},c}^{(i)}\right)$. In terms of calculation, the correlation factors $\rho_{f\mathbb{1}}^{(i)}$ and $\rho_{f\mathbb{1},c}^{(i)}$ can be estimated as

$$\begin{aligned}
\hat{\rho}_{f\mathbb{1}}^{(i)} &= \frac{1}{N_s - 1} \sum_{t=1}^{N_s-1} \frac{\frac{1}{N_c}\sum_{j=1}^{N_c} f_i^{(j,t)} \mathbb{1}_i^{(j,t)} - \hat{h}_i \hat{p}_c^{(i)}}{\sqrt{\frac{1}{N_c}\sum_{j=1}^{N_c} f_i^{(j,t)2} - \hat{h}_i^2} \sqrt{\frac{1}{N_c}\sum_{j=1}^{N_c} \mathbb{1}_i^{(j,t)2} - \hat{p}_c^{(i)2}}} \\
\hat{\rho}_{f\mathbb{1},c}^{(i)} &= \frac{1}{N_s - 1} \sum_{t=1}^{N_s-1} \frac{\frac{1}{2N_c}\sum_{j=1}^{N_c} \left[f_i^{(j,t)} \mathbb{1}_i^{(j,t+1)} + f_i^{(j,t+1)} \mathbb{1}_i^{(j,t)}\right] - \hat{h}_i \hat{p}_c^{(i)}}{\sqrt{\frac{1}{N_c}\sum_{j=1}^{N_c} f_i^{(j,t)2} - \hat{h}_i^2} \sqrt{\frac{1}{N_c}\sum_{j=1}^{N_c} \mathbb{1}_i^{(j,t)2} - \hat{p}_c^{(i)2}}}
\end{aligned} \tag{A. 10}$$

## Appendix B. Performance metrics for BUS/aBUS

In BUS and aBUS, the evidence estimator is given as $\hat{Z} = c^{-1}\hat{P}_f$, where the constant $c^{-1}$ should be chosen as the maximum of the likelihood function $L(\boldsymbol{\theta})$, and the "failure" probability estimator $\hat{P}_f$ is given as

$$\hat{P}_f = \Pr[g(\boldsymbol{\theta}, \pi) \leq 0] = \prod_{i=0}^{M} \underbrace{\frac{1}{N} \sum_{k=1}^{N} \mathbb{1}\left[g(\boldsymbol{\theta}_k^{(i)}, \pi) \leq h_{i+1}\right]}_{\hat{p}_c^{(i)}} \tag{B. 1}$$

Here, the limit state function is expressed as $g(\boldsymbol{\theta}, \pi) = \pi - cL(\boldsymbol{\theta})$, with $\pi$ being the standard uniform RV. The thresholds $h_i$ are adaptively selected in SuS and satisfy $+\infty = h_0 \geq h_1 \geq \cdots h_i \geq h_i \geq \cdots \geq h_M = 0$.

Since the maximum of the likelihood function is unknown a priori, $c^{-1}$ should be estimated adaptively and thus subject to uncertainty. Considering its small uncertainty compared to that of the estimator $\hat{P}_f$, we ignore the uncertainty in estimating $c^{-1}$, and approximate the c.o.v of $\hat{Z}$ as

$$\delta_z^2 \approx \delta_{p_f}^2 \approx \sum_{i=0}^{M} \delta_p^{(i)^2} \tag{B.2}$$

where $\delta_{p_f}$ and $\delta_p^{(i)}$ are the c.o.v. of $\hat{P}_f$ and $\hat{P}_c^{(i)}$, respectively. We adopt the formula proposed in Ref. [28] to obtain an estimated value of $\delta_p^{(i)}$ in calculation.

The posterior samples obtained by BUS/aBUS are those generated in the last iteration corresponding to $h_M = 0$. Therefore, the ESS of BUS/aBUS can be easily estimated as

$$N_{ess} = N \frac{\text{VAR}[\check{Z}]}{\text{VAR}[\hat{Z}]} \tag{B.3}$$

where $N$ represents the sample size in the last iteration, $\text{VAR}[\hat{Z}]$ and $\text{VAR}[\check{Z}]$ are respectively the variance of evidence estimator of BUS/aBUS considering and not considering the correlation in the parallel MCMC sampling.